\documentclass[reprint,onecolumn,groupedaddress,showpacs,notitlepage,longbibliography]{revtex4-1}

\usepackage{graphicx}
\usepackage{caption}
\usepackage{subcaption}
\usepackage{color}
\usepackage{amsmath}
\usepackage{amssymb}
\usepackage{verbatim}
\usepackage[english]{babel}
\usepackage{hyperref}

\DeclareGraphicsExtensions{.pdf,.png}

\begin{document}

\title{Model-free measure of coupling from embedding principle}

\author{Chetan Nichkawde}

\email{chetan.nichkawde@mq.edu.au}

\affiliation{Department of Physics and Astronomy, Macquarie University, Sydney,
Australia}

\date{\today}

\pacs{05.45.Tp, 89.65.Gh, 05.45.-a, 05.45.Xt}

\begin{abstract}
A model-free measure of coupling between dynamical variables is built from time series embedding principle.
The approach described does not require a mathematical form for the dynamics to be assumed.
The approach also does not require density estimation which is an intractable problem in high dimensions.
The measure has strict asymptotic bounds and is robust to noise. 
The proposed approach is used to demonstrate coupling between complex time series from the finance world. 
\end{abstract}

\maketitle

Probing for coupling between dynamical variables is a problem of fundamental
interest in a variety of disciplines. In real world settings the
underlying model is often unknown and we only have time series measurements.
An information-theoretic approach seeks to assess coupling between
two time series measurements by ascertaining if additional information
about the future state of a variable can be gained by including the
second variable in a discriminative model. This measure which is popularly
termed \textit{transfer entropy} finds the difference between conditional
entropy of future states between models that contain and do not contain
the second variable\cite{schreiber2000measuring}. The notion of determining
coupling via a predictability approach also forms the basis of Granger
causality which addresses the same question by assessing the predictability
of one of the variables using linear autoregressive models in which
the second variable is present or absent\cite{granger1969investigating}.
The Granger approach has been generalized to the nonlinear case by
using a nonlinearly transformed feature space\cite{marinazzo2008kernel}.
Transfer entropy and Granger causality were shown to be equivalent
for Gaussian variables\cite{barnett2009granger}. While determining
coupling via the predictability mechanism notionally appears to be
true, the idea has never been mathematically established, both in transfer
entropy and Granger causality. Moreover, these measures do not have
any asymptotic limit\cite{kaiser2002information}. A zero transfer
entropy in one direction must be obtained in order to conclude directionality\cite{kaiser2002information}.
However, the measure depends on accurate evaluation of conditional
densities which is often obtained by expressing those in terms of
joint probabilities. The density evaluation suffers from bias and
in practice one obtains a nonzero value of transfer entropy even for
cases where it is theoretically supposed to be zero\cite{kaiser2002information,Staniek}.
Staniek et al\cite{Staniek} proposed symbolic quantization using
ordinal patterns for density estimation. The directionality of coupling
for example in this case was ascertained by comparing two values for
transfer entropy where the driving variables are switched. The absolute
value of these entropies however cannot be compared with another case.
Transfer entropy calculation returns a nonzero value due to statistical
bias in density estimation both for uncoupled and fully synchronized
variables\cite{Staniek}. This makes it difficult to distinguish between
these two cases. In both the approaches above, a very important parameter
is the memory or Markov order of the underlying dynamical process.
This parameter is chosen in an \textit{ad hoc} manner. Some processes
can have long memory which warrants that even a model-free measure
of transfer entropy must ascertain these high dimensional densities.
This is especially true for deterministic complex signals such as
those found in a chaotic system.

The above highlighted issues are addressed in this paper by presenting
a unified framework for coupling detection. Following are the salient
features of the proposed framework: 
\begin{enumerate}
\item A model-free measure of Markov order is first used in determining
the memory of the dynamical system. 
\item The measure of coupling is built from state space reconstruction based
on first principles. A necessary condition for coupling is first established.
A measure is then built to assess this necessary condition. 
\item This measure is convergent to an asymptotic absolute limit both for
uncoupled and fully coupled cases. The lower bound is zero for completely
uncoupled variables and the upper bound is one for fully coupled variables.
The proposed approach is the best way to avoid false positives owing
to a strict lower bound which is not affected by the amount of noise
in the signal. Detection of true positives is also robust to the presence
of a large amount of noise. 
\item The approach is model-free and therefore is not limited by the assumptions
of a parametric model. 
\item The approach can distinguish between uncoupled and fully synchronized
systems. Fully synchronized systems can be detected with the measure
taking a value of one. 
\end{enumerate}

The outline of this paper is as follows: Section~\ref{sec:ncon}
establishes the necessary condition for two variables to be coupled
from attractor reconstruction based first principles. Section~\ref{sec:stat}
elaborates a statistical measure to assess this necessary condition.
This completes the foundation for this work. Section~\ref{sec:order}
explains an approach to determine the Markov order of a time series.
Section~\ref{sec:mdop} discusses the choice of appropriate time series embedding for the methodology described in this paper. 
Section~\ref{sec:noise} introduces the procedure
by probing for coupling between $x$ and $y$ variables of Rossler
system. Robustness of statistics to noise is also demonstrated. Section~\ref{sec:directionality}
conjectures that the proposed approach can also be used to ascertain
the directionality of coupling and this is demonstrated with the help
of asymmetrically coupled Lorenz system. The effectiveness of the
approach to distinguish between uncoupled and fully synchronized systems
is also discussed. In Section~\ref{sec:realworld}, the application
of this method is demonstrated on an example from the finance world. 
It is shown that the currency exchange rate Canadian Dollar-Japanese Yen and oil prices
are strongly coupled. Section~\ref{sec:conclusions} concludes the
findings of this paper.

\section{\label{sec:ncon}Necessary condition for coupling}

Consider a dynamical system 
\begin{align}
\dot{\mathbf{x}}=f\left(\mathbf{x}\right)
\end{align}
 where $\mathbf{x}\in\mathbb{R}^{n}$ is the state vector of the system.
Let $x_{1},~x_{2},...,x_{n}$ be the state variables. Let $\mathbb{M}$
be the manifold on which the states of the system asymptotes as it
evolves over time. This manifold is termed as the attractor of the
system. For an attractor manifold $\mathbb{M}$ with $\mathbf{x}(t)$
as the state of the system at time $t$ and flow $f:\mathbb{M}\to\mathbb{M}$,
the manifold defined by the $F(h,f,\tau_{1},...,\tau_{m}):\mathbb{M}\to\mathbb{R}^{m}$
is generically an \textit{embedding} for 
\begin{align}
F(\mathbf{x}(t))= & [x_{m}(t),x_{m}(t-\tau_{1}),x_{m}(t-\tau_{2}),\nonumber \\
 & ...,x_{m}(t-\tau_{m})]\label{eq:embedding}
\end{align}
 if $m\geq2d+1$ where $d$ is the dimension of the original state
space\cite{Whitney1936,Packard,Takens,Sauer}. $x_{m}$ above is one
of the observed state variables. This \textit{embedding} is a diffeomorphic
map between the attractor and the reconstructed state space. Consider
two different embeddings of $\mathbb{M}$, $F_{1}:\mathbb{M}\to\mathbb{R}^{m_{1}}$
and $F_{2}:\mathbb{M}\to\mathbb{R}^{m_{2}}$ formed by delay embedding
procedure given by Eq.~(\ref{eq:embedding}). Let $\mathbf{y}\to\mathbb{R}^{m_{1}}$
and $\mathbf{z}\to\mathbb{R}^{m_{2}}$ be vectors in transformed coordinates
for $F_{1}$ and $F_{2}$. These can be expressed in terms of time
delay variables as: 
\begin{align}
\mathbf{y} & =[y(t),y(t-\tau_{y_{1}}),..,y(t-\tau_{y_{m_{1}}})]\\
\mathbf{z} & =[z(t),z(t-\tau_{z_{1}}),..,z(t-\tau_{z_{m_{1}}})]
\end{align}
 where $y(t)$ and $z(t)$ are two of the state variables for $\mathbf{x}$.
Thus, 
\begin{align*}
\left[\begin{array}{c}
y(t)\\
y(t-\tau_{y_{1}})\\
.\\
.\\
y(t-\tau_{y_{m_{1}}})
\end{array}\right]=F_{1}\left(\begin{array}{c}
x_{1}\\
x_{2}\\
.\\
.\\
x_{n}
\end{array}\right)
\end{align*}
 and 
\begin{align*}
\left[\begin{array}{c}
z(t)\\
z(t-\tau_{z_{1}})\\
.\\
.\\
z(t-\tau_{z_{m_{2}}})
\end{array}\right]=F_{2}\left(\begin{array}{c}
x_{1}\\
x_{2}\\
.\\
.\\
x_{n}
\end{array}\right)~.
\end{align*}
 This can be written in compact manner as: 
\begin{align}
\mathbf{y}=F_{1}(\mathbf{x})\label{eq:emy}
\end{align}
 and 
\begin{align}
\mathbf{z}=F_{2}(\mathbf{x})~.\label{eq:emz}
\end{align}
 $F_{1}$ and $F_{2}$ are continuous. $F_{1}^{-1}$ and $F_{2}^{-1}$
also exist and are continuous. These are the properties of diffeomorphic
maps such as $F_{1}$ and $F_{2}$.

\begin{figure}
\includegraphics[width=\textwidth]{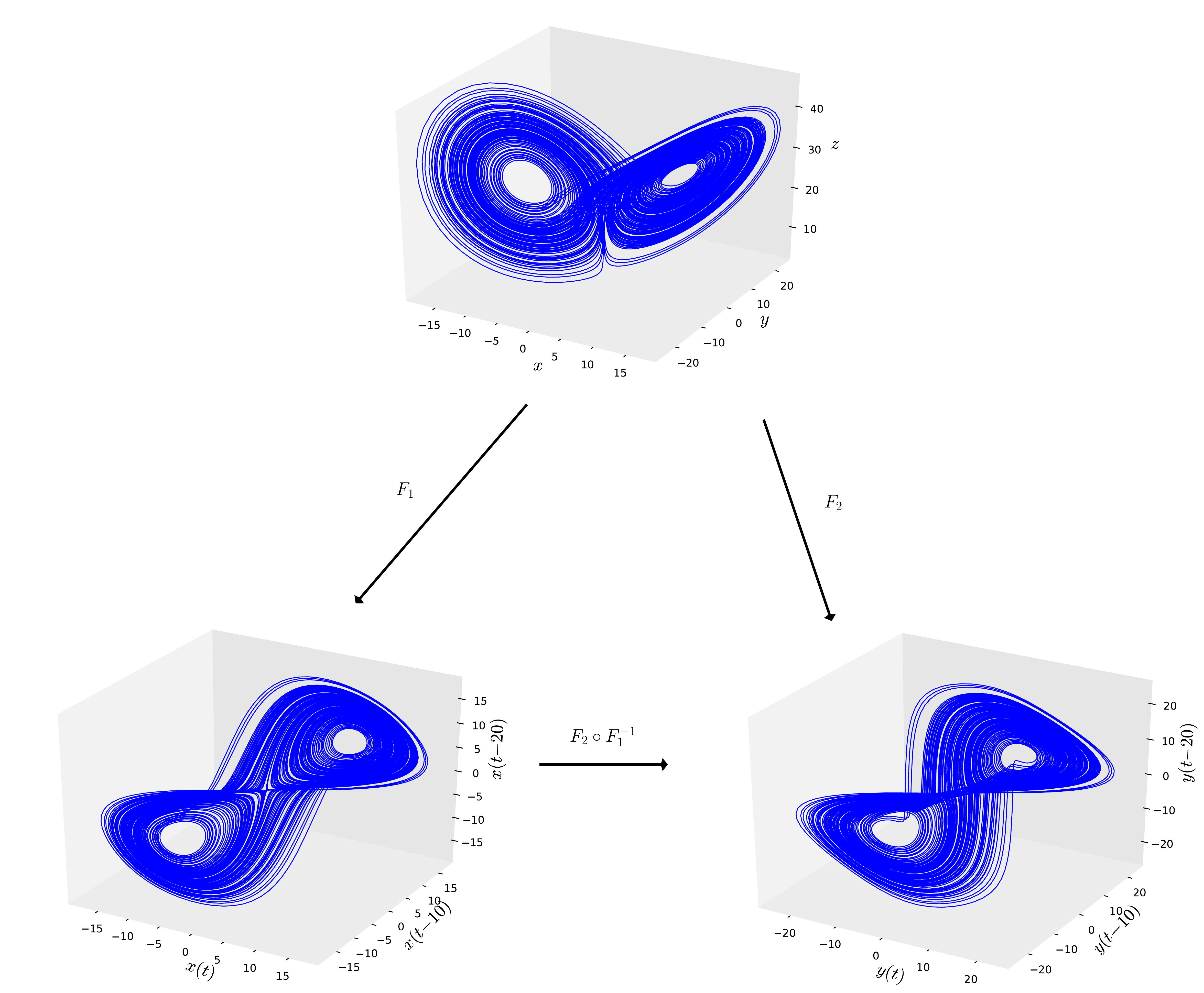} \caption{\label{fig:lorenzcouple}$x$ and $y$ variables of the Lorenz systems
are coupled. A continuous map $F_{1}$ exists between original attractor
and the attractor reconstructed using time delay embedding of $x$
variable. Similarly, a continuous map $F_{2}$ exists for $y$ variable.
Thus, coupling between $x$ and $y$ implies that a continuous map $F_{2}\circ F_{1}^{-1}$
exists between embeddings of $x$ and $y$.}
\end{figure}

Two variables $y$ and $z$ are coupled if they belong to the same
dynamical system. With this notion of coupling in mind, let us consider
time delay embeddings of $y$ and $z$ given by Eqs.~(\ref{eq:emy})
and (\ref{eq:emz}). $\mathbf{x}$ is expressed in terms of $\mathbf{y}$
in Eq.~(\ref{eq:emy}) as 
\begin{align}
\mathbf{x}=F_{1}^{-1}(\mathbf{y})~.\label{eq:xz}
\end{align}
 Putting the above in Eq.~(\ref{eq:emz}) yields: 
\begin{align}
\mathbf{z}=(F_{2}\circ F_{1}^{-1})(\mathbf{y})~.
\end{align}
 Since $F_{2}$ and $F_{1}^{-1}$ are continuous, their composition
$F_{2}\circ F_{1}^{-1}$ is also continuous. Please see Appendix~\ref{sec:CompositionProof}
for proof of this. This implies that if $y$ and $z$ belong to the
same dynamical system, their exists a continuous function map between
their suitable time delay embeddings. Thus, 
\begin{align}
\mathbf{z}=\Psi(\mathbf{y})
\end{align}
 where $\Psi=F_{2}\circ F_{1}^{-1}$ is continuous. It is easy to
see that $y$ and $z$ above are inter-changeable. The concept is
illustrated in Fig.~\ref{fig:lorenzcouple} which shows the Lorenz
attractor along with time delay embeddings using $x$ and $y$ variables
of this system. A continuous functional map $F_{1}$ exists between
the original attractor and the time delay embedding with the $x$ variable.
Similarly, a continuous $F_{2}$ exists for the embedding with the $y$
variable. This implies that a continuous map $F_{2}\circ F_{1}^{-1}$
also exists between embeddings of $x$ and $y$.

\section{\label{sec:stat}Statistics to assess the necessary condition}

Consider a point $\mathbf{y}_{c}$ on the reconstructed manifold for
variable $y$. Let this point be mapped to $\mathbf{z}_{c}$ by $\Psi$.
If $\Psi$ is continuous at the point $\mathbf{y}_{c}$ then for every
$\epsilon>0$ there exists a $\delta>0$ such that for all $\mathbf{y}$:
\begin{align*}
\left|\mathbf{y}-\mathbf{y}_{c}\right| & <\delta\Rightarrow\left|\Psi(\mathbf{y})-\Psi(\mathbf{y}_{c})\right|<\epsilon
\end{align*}
 or 
\begin{align}
\left|\mathbf{y}-\mathbf{y}_{c}\right| & <\delta\Rightarrow\left|\mathbf{z}-\mathbf{z}_{c}\right|<\epsilon~.
\end{align}

Consider two time series of $y(t)$ and $z(t)$ of equal length $N$.
Consider embeddings of $\mathbf{y},~\mathbf{z}$ of $y(t)$ and $z(t)$
respectively. Let their exist a map $\mathbf{z}=\Psi(\mathbf{y})$
between $\mathbf{y}$ and $\mathbf{z}$. Consider a point $\mathbf{y}_{c}$
on the embedding $\mathbf{y}$. Let this point be mapped to $\mathbf{z}_{c}$
by $\Psi$. Consider a ball of size $\epsilon$ centered around $\mathbf{z_{c}}$.
Let their be $n_{\epsilon}$ number of points inside this ball. Consider
a ball of minimum size $\delta$ centered around $\mathbf{y_{c}}$
such that all points within this ball are mapped by $\Psi$ to points within
the $\epsilon$ ball. $\delta$ can be found by starting from small
values and gradually increasing it until the above condition is satisfied.
Let there be $n_{\delta}$ number of points within the $\delta$ ball.

The probability $p_{\epsilon}$ of point from $\mathbf{y}$ space
being mapped into the $\epsilon$ ball by random chance is 
\begin{align}
p_{\epsilon}=\frac{n_{\epsilon}}{N}~.
\end{align}
 The probability of $k$ successful draws out of $n_{\delta}$ points
drawn from a sample of size $N$ with probability of success for each
draw being $p_{\epsilon}$ is given by the binomial distribution:
\begin{align}
p(k)=\frac{n_{\delta}!}{k!(n_{\delta}-k)~!}p_{\epsilon}^{k}\left(1-p_{\epsilon}\right)^{n_{\delta}-k}~.\label{eq:binom}
\end{align}
 This is also the probability of getting $k$ number of heads out
of $n_{\delta}$ tosses of a coin where the probability of getting
a head in each toss is $p_{\epsilon}$. Our null hypothesis for continuity
is that $n_{\delta}$ points inside the $\delta$ are mapped into
the $\epsilon$ ball by random chance\cite{Pecora1}. The probability
$p_{\delta}$ of all $n_{\delta}$ points inside the $\delta$ ball
landing into the $\epsilon$ ball by random chance is 
\begin{align}
p_{\delta}=p_{\epsilon}^{n_{\delta}}~.
\end{align}
 This event lies in the tail of the distribution given by Eq.~(\ref{eq:binom}).
The above value should be small relative to the maximum probability
of such an event, $p_{max}$, in order to reject the null hypothesis.
The likelihood of this event happening is defined as $\frac{p_{\delta}}{p_{max}}$
where 
\begin{align}
p_{max}=\operatorname*{arg\, max}_{k}p(k)~.
\end{align}
 Thus, the statistic for continuity is thus defined as: 
\begin{align}
\theta(\epsilon,\mathbf{y}_{c})=1-\frac{p_{\delta}}{p_{max}}~.
\end{align}
$\theta(\epsilon,\mathbf{y}_{c})$ is bounded below by 0 and bounded above by 1. 
If the value of $\theta(\epsilon,\mathbf{y}_{c})$ is high, then
the function is continuous at $\mathbf{y}_{c}$. Let $d$ be the length
of the diagonal of the bounding box for the $\mathbf{z}$ attractor.
$\epsilon$ can then be expressed in terms of a fraction of $d$ as
\begin{align}
\epsilon=\epsilon_{f}d
\end{align}
 where $0\le\epsilon_{f}\le1$. The value of $\theta$ is averaged
over all the sample points and can be expressed as 
\begin{align}
\theta_{avg}(\epsilon_{f})=\frac{1}{N}\sum_{i=1}^{N}\theta(\epsilon_{f},\mathbf{y}_{i})~.
\end{align}
 $\theta_{avg}(\epsilon_{f})$ above represents the measure of coupling
being proposed in this paper and will be referred to as the \emph{coupling statistic} in the remainder of this paper.

\section{\label{sec:order}Forward causality: Determining the Markov order}

A time series $x_{t}$ $(t=1,2,...,N)$ is an $n^{th}$ order Markov
process if the probability of $x_{t+1}$, conditioned on all the previous
values of $x$ in time, is independent of $x_{t-m-1}$ where $m>n$.
This can written as 
\begin{align}
p(x_{t+1}|x_{t},x_{t-1},...,x_{1})=p(x_{t+1}|x_{t},x_{t-1},...,x_{t-n-1})~.
\end{align}
 $x_{t-m-1}$ where $m>n$ are irrelevant in determining the transition
probability. The same has been termed as \emph{irrelevancy} in Ref.~\cite{Casdagli}.
Irrelevancy is an important concept and is fundamental to time series
modeling but has received very little attention in the literature.
Causality for all practical purposes is lost beyond a certain time
in the past. Events beyond a time limit in the past do not have any
bearing on the future states.

The Markov order is determined by examining how well conditioned the
future states are given a certain Markov order $n$. With a proper
choice of Markov order $n$, the variance of future values of $x$,
conditioned on the present, would be minimized. The variance of $x(t+T)$
for a small sized ball $B_{r}(\mathbf{x}(t))$ of radius $r$
around $\mathbf{x}\in R^{n}$, normalized by the size of the ball,
is given by 
\begin{align}
\sigma_{r}^{2}(T,\mathbf{x}(t))=\frac{1}{r^{2}}Var(x(t+T)|B_{r}(\mathbf{x}(t)))~.\label{eq:sigr}
\end{align}
 Uzal et al\cite{Uzal} modified the above by taking the integral of the
$\sigma_{r}^{2}(T,\mathbf{x}(t))$ from zero to prediction horizon $T_{M}$:
\begin{align}
\sigma_{r}^{2}(\mathbf{x})=\frac{1}{T_{M}}\int_{0}^{T_{M}}\sigma_{r}^{2}(T,\mathbf{x})dT~.
\end{align}
 Assuming a certain Markov order $m$, $k$ nearest neighbors of $\mathbf{x}(t)$
are considered. Let us denote this set, which includes $\mathbf{x}(t)$
itself, as $U_{k}(\mathbf{x}(t))$. The conditional variance of $x$
at $t+T$ is approximated using these nearest neighbors as 
\begin{align}
E_{k}^{2}(T,\mathbf{x}(t))=\frac{1}{k+1}\operatorname*\sum_{x'\in U_{k}(\mathbf{x}(t))}\left[x'(T)-u_{k}(T,\mathbf{x}(t))\right]^{2}\label{eq:Ek}~,
\end{align}
 where $x'(T)$ is the future value of $x$ corresponding to $\mathbf{x'}$,
and 
\begin{align}
u_{k}(T,\mathbf{x}(t))=\frac{1}{k+1}\operatorname*\sum_{x'\in U_{k}(\mathbf{x}(t))}x'(T)~.
\end{align}
 The expression in Eq.~(\ref{eq:Ek}) is averaged up to a prediction
horizon $T_{M}$. $E_{k}(\mathbf{x}(t))$ can then be defined without
explicit dependence on $T$ as 
\begin{align}
E_{k}^{2}(\mathbf{x}(t))=\frac{1}{p}\operatorname*\sum_{j=1}^{p}E_{k}^{2}(T_{j},\mathbf{x}(t))~,
\end{align}
 where the sum is over $p$ sampled times $T_{j}$ in $[0,T_{M}]$.
The size of the neighborhood for conditional variance estimation is
given by 
\begin{align}
\epsilon_{k}^{2}(\mathbf{x})=\frac{2}{k(k+1)}\operatorname*{\operatorname*\sum_{\mathbf{x'},\mathbf{x''}\in U_{k}(\mathbf{x}(t))}}_{\mathbf{x''}\ne\mathbf{x}}\|\mathbf{x'}-\mathbf{x''}\|^{2}~.
\end{align}
 This is a measure of the characteristic radius of $U_{k}(\mathbf{x}(t))$.
The noise amplification which is a measure of conditional variance
is the given as 
\begin{align}
\sigma_{k}^{2}(\mathbf{x})=\frac{E_{k}^{2}(\mathbf{x})}{\epsilon_{k}^{2}(\mathbf{x})}~.
\end{align}
 This is averaged over $N$ points as 
\begin{align}
L_{n}=\log\left(\frac{1}{N}\operatorname*{\sum}_{i=1}^{N}\sigma_{k}^{2}(\mathbf{x_{i}})\right)~.\label{eq:sigma}
\end{align}
 The minimum of the above function with various considered values
$n$, would be an appropriate choice for Markov order of the time
series. The variation of $L_{n}$ with $n$ will be shown later for
an economic time series example in Section~\ref{sec:realworld}.

\section{\label{sec:mdop}Choosing an appropriate embedding}

Once the memory of the system is approximated using the procedure
elaborated in the previous section, the time series can be embedded
onto a much lower dimensional manifold. Markov approximation for deterministic
nonlinear dynamics have a strongly perforated structure and the dynamics
can be appropriately represented by only a few of the time delay co-ordinates\cite{Holstein}.
Nearest neighbor searches in dimension greater than 20 can only be done
$\mathcal{O}(N^{2})$. A minimal embedding procedure can allow the use
of fast search-tree-based methods for the nearest neighbors search.
These methods perform a nearest neighbor search operations in $\mathcal{O}(N\log N)$.
However, if the data is very noisy and the Markov order is low enough($\mathtt{\sim}\mathcal{O}(20)$),
then an embedding with all the delays can be used. The nearest neighbor
rank becomes more robust to noise with increasing embedding dimension.
It was shown in Ref.~\cite{hegger2001circuits} that the nearest
neighbor distance in noisy and noise-free cases are related by the
following relationship for a sufficiently high embedding dimension
$m$: 
\begin{align*}
d_{noisy}^{2}\approx d_{clean}^{2}+2m\xi^{2}~,
\end{align*}
 where $d_{noisy}$ is the distance in the noisy case, $d_{clean}$
is the distance in the noise-free case and $\xi^2$ is the noise variance.
Since, the statistics used in this paper depends only on the nearest
neighbor ranks and not the actual nearest neighbor distances, it is
advisable to use as high an embedding dimension as possible. For a
high Markov order case, an approach to minimally embedding the time
series on a low dimensional manifold, as described in Ref.~\cite{Nichkawde},
can be used. This methodology recursively chooses delays that maximize
derivatives on the project manifold. The objective functional is of
the following form: 
\begin{align}
\log\left[\beta_{d}(\tau)\right]=\left<\log\phi'_{d_{ij}}\right>~.\label{eq:mdop}
\end{align}
 In the above equation, $\phi'_{d_{ij}}$ is the value of the directional
derivative evaluated in the direction from the $i^{th}$ to the $j^{th}$
point of the projected attractor manifold which happens to be the
nearest neighbor. The recursive optimization of objective functional
given by Eq.~(\ref{eq:mdop}) eliminates the largest number of false
nearest neighbors between successive reconstruction cycles and thus
helps achieve an optimal minimal embedding\cite{Nichkawde}. 
This procedure would be referred to as MDOP (maximising derivatives
on projected manifold). The difference obtained between choosing a
minimal embedding versus choosing an embedding with a delay of one
and embedding dimension equal to the Markov order will be reported
in Section~\ref{sec:realworld}.

\begin{figure*}
    \centering
    \begin{subfigure}[t]{0.45\textwidth}
        \centering
        \includegraphics[width=\textwidth]{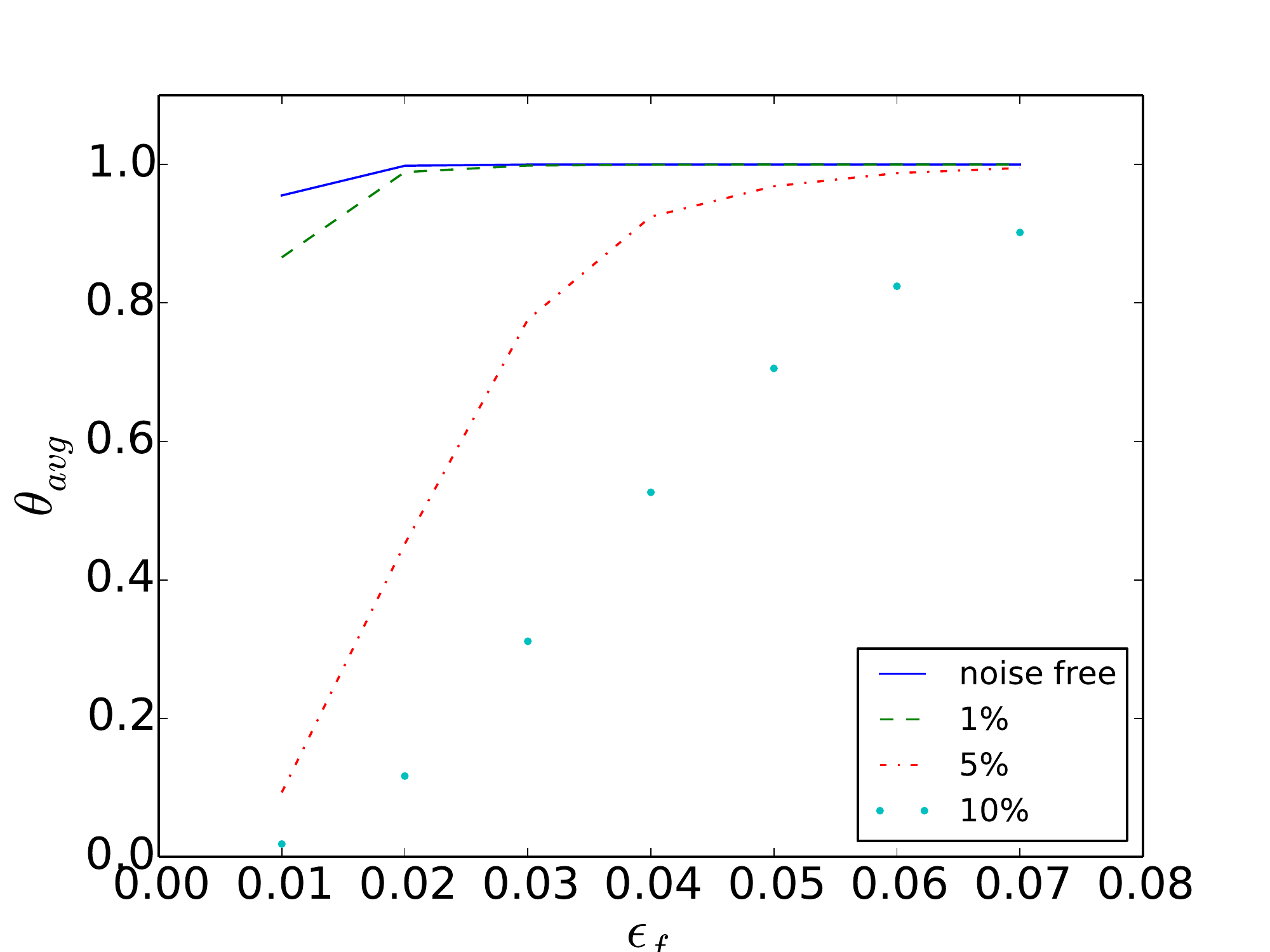}
    \caption{$\theta_{avg}(\epsilon_{f})$ for probing
coupling between $x$ and $y$ variables of Rossler system for noise
free case and with noise levels of 1\%, 5\% and 10\%. Coupling
can be detected even with 10\% Gaussian noise.}
    \label{fig:rossler_cstat}
    \end{subfigure}
    \begin{subfigure}[t]{0.45\textwidth}
        \centering
        \includegraphics[width=\textwidth]{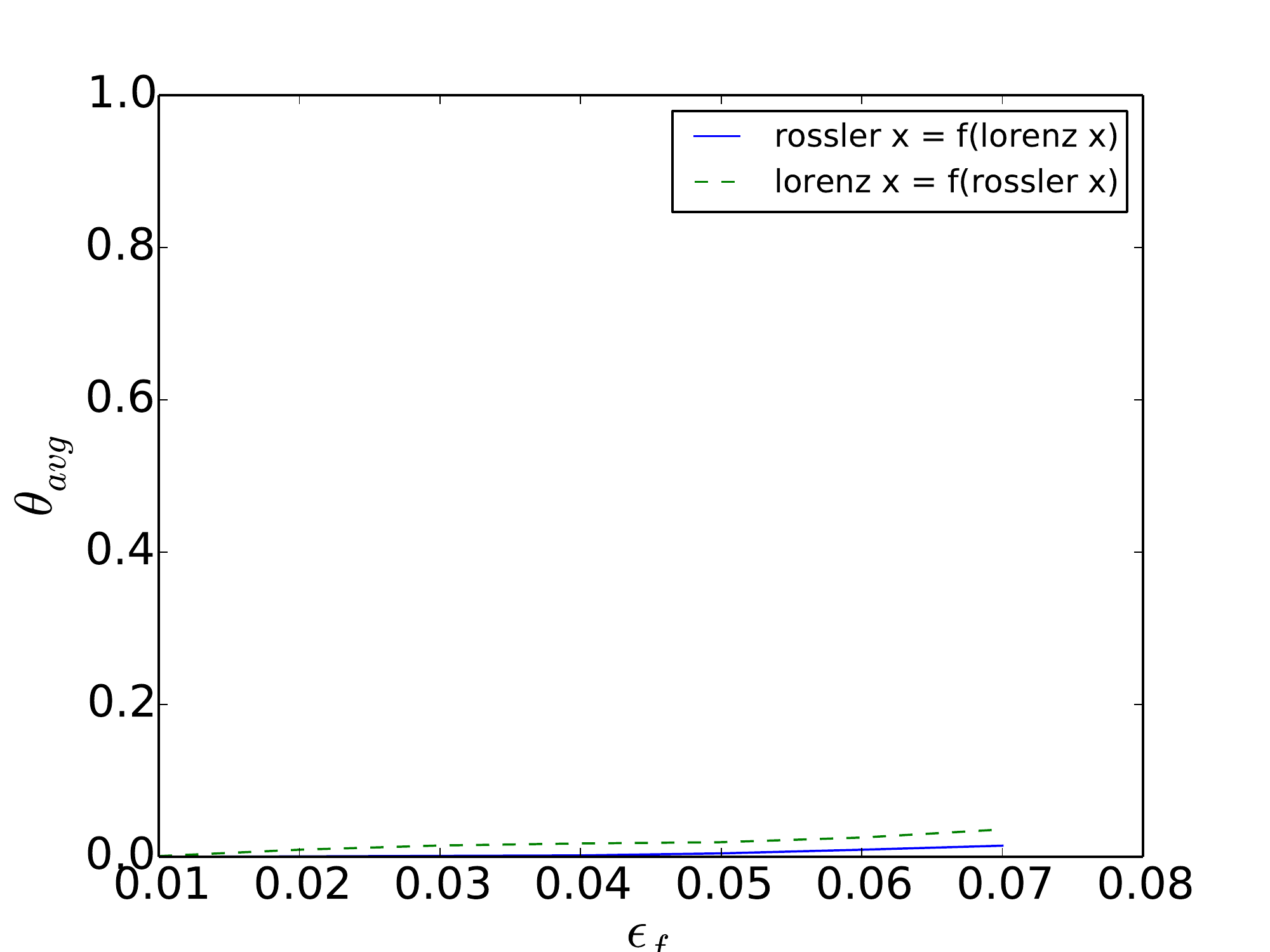} 
        \caption{$\theta_{avg}(\epsilon_{f})$ for probing
coupling between Rossler $x$ and Lorenz $x$.}
        \label{fig:rosslerlorenzcstat}
    \end{subfigure}
    \caption{Coupling statistics for noisy coupled and uncoupled time series.}
\end{figure*}
\setcounter{subfigure}{0}
\begin{figure*}
    \centering
    \begin{subfigure}[t]{0.45\textwidth}
        \centering
        \includegraphics[width=\textwidth]{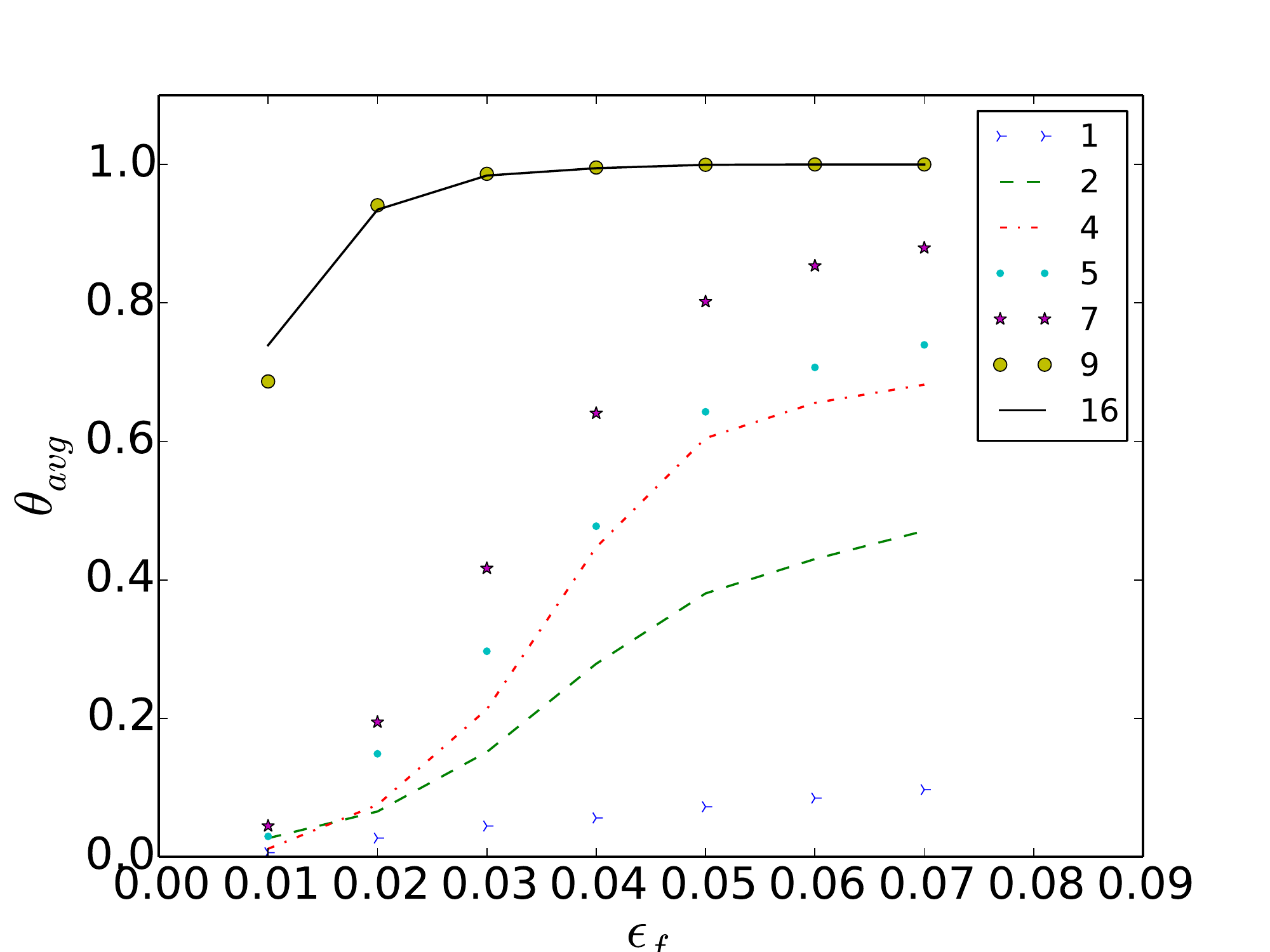}
        \caption{$\theta_{avg}(\epsilon_{f})$
for $\mathbf{x_{1}}=f_{1}(\mathbf{x_{2}})$ for $\alpha$ of 1, 2,
4, 5, 7, 9 and 16. Mild coupling for $\alpha$ of 1, 2, 4, 5 and 7.
The variables become synchronized and fully coupled for $\alpha$
of 9 and 16.}
    \label{fig:coupledlorenz_x1x2_cstat}
    \end{subfigure}
    \begin{subfigure}[t]{0.45\textwidth}
        \centering
        \includegraphics[width=\textwidth]{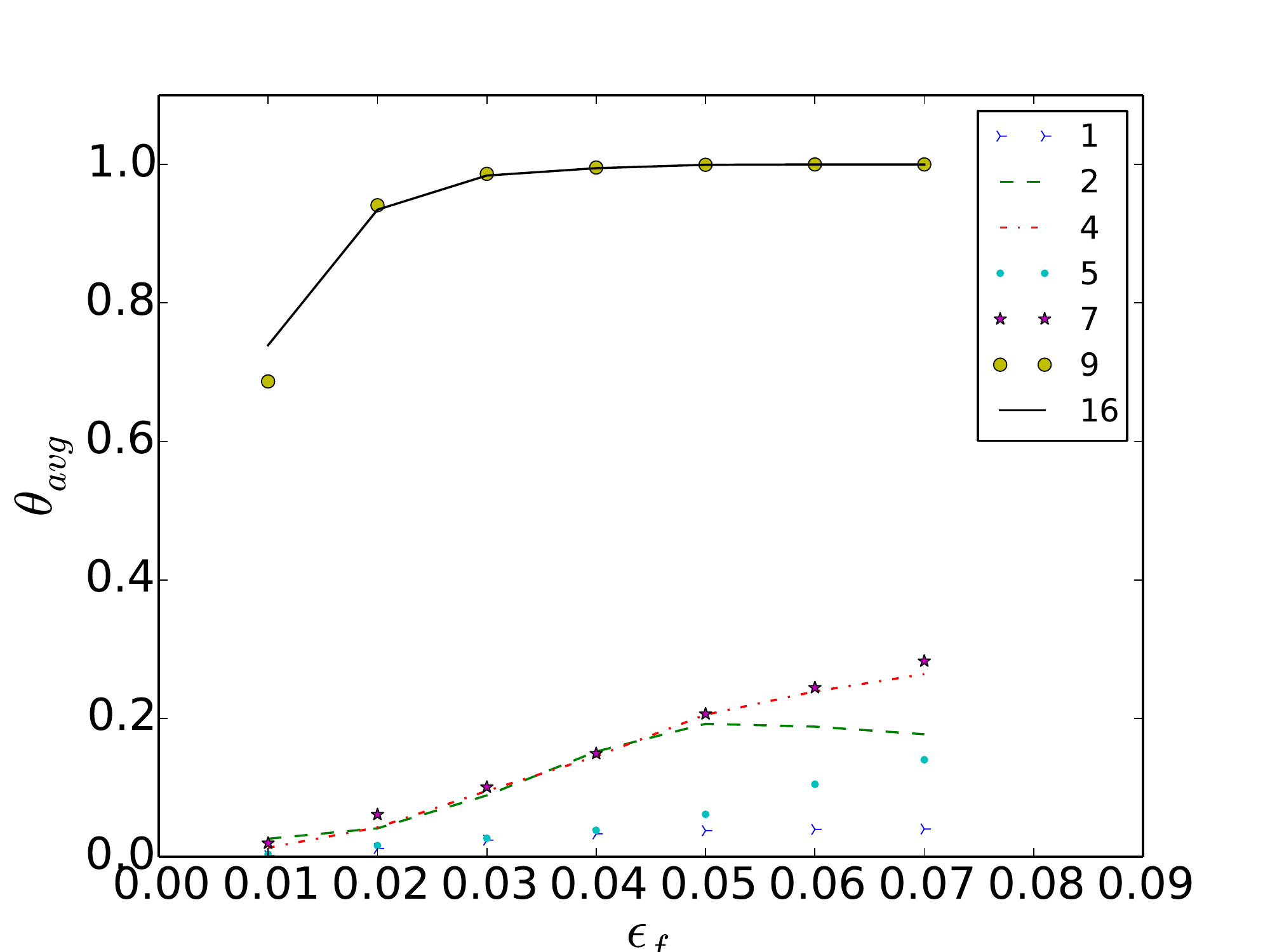}
\caption{$\theta_{avg}(\epsilon_{f})$
for $\mathbf{x_{2}}=f_{2}(\mathbf{x_{1}})$ for $\alpha$ of 1, 2,
4, 5, 7, 9 and 16. $x_{2}$ drives $x_{1}$ and thus the coupling
is unidirectional. The values of $\theta_{avg}(\epsilon_{f})$ is
small for $\alpha$ of 1, 2, 4, 5 and 7 as there is no drive from
$x_{1}$ to $x_{2}$. The variables however become synchronized for
$\alpha$ of 9 and 16 which is manifest as values of $\theta_{avg}(\epsilon_{f})$
close to one. Thus the approach proposed can distinguish between uncoupled
and fully synchronized systems.}
    \label{fig:coupledlorenz_x2x1_cstat}
    \end{subfigure}
    \caption{Coupling between unidirectionally coupled Lorenz systems.}
\end{figure*}
\setcounter{subfigure}{0}

\section{\label{sec:noise}Coupling statistics for coupled and uncoupled systems}
Let us consider the $x$ and $y$ variables of the Rossler system\cite{Rossler}.
This system is modeled by a set of 3 coupled ordinary differential equations
\begin{align*}
\dot{x} & =-y-z\\
\dot{y} & =x+ay\\
\dot{z} & =b+z(x-c)
\end{align*}
 with parameter values $a=0.2,~b=0.2,~c=5.7$. 10000 points were sampled
with a $\delta t$ of 0.05 with initial condition $[0.1,0,0]$. Time
delays for embedding are determined using the method prescribed in
Ref.~\cite{Nichkawde}. Delays of 30 and 17 are found to be most optimal
for the $x$ variable whereas delays of 31 and 17 are found to be most optimal
for the $y$ variable. 
$\theta_{avg}$ is evaluated for values of $\epsilon_{f}$ ranging
from 0.01 to 0.07. These values are shown in Fig.~\ref{fig:rossler_cstat}
for noise free case and with noise levels of 1\%, 5\% and 10\%.
The value $\theta_{avg}$ converges quickly to the expected value
of 1 for very small $\epsilon_{f}$ for noise-free and 1\% noise cases.
Further increase in noise levels degrades the statistics. 
The coupling can still be easily ascertained for noise levels upto 10\%.

The behavior of coupling statistics for two uncoupled systems is shown in Fig.~\ref{fig:rosslerlorenzcstat}.
One of the variables is chosen as $x$ variable for Rossler system and is taken same as in the previous
example. A time series of equal length is generated for Lorenz system
which is given by the following set of equations: 
\begin{align*}
\dot{x} & =\sigma(y-x)\\
\dot{y} & =-xz+rx-y\\
\dot{z} & =xy-\beta z
\end{align*}
 where $\sigma=10,~r=28,~\beta=8.0/3$.
The $x$ variable for this system is taken as the second variable. 
The value of $\theta_{avg}$ for values of $\epsilon_{f}$ ranging
from 0.01 to 0.07 are shown in Fig.~\ref{fig:rosslerlorenzcstat}.
Very low values of the measure is indicative of no coupling between
the variables, as expected. 
The measure is exactly zero when two noise signals are probed for coupling. 

\section{\label{sec:directionality}Directionality and synchronization}

The formulation proposed in this paper says nothing about the directionality
of the coupling. Nevertheless, the following is still conjectured:
Two variables $x_{1}$ and $x_{2}$ are unidirectionally coupled with
$x_{2}$ driving $x_{1}$ if the map $\mathbf{x_{1}}=f_{1}(\mathbf{x_{2}})$
between embedding $\mathbf{x_{1}}$ of $x_{1}$ and $\mathbf{x_{2}}$
of $x_{2}$ is continuous, whereas, the map $\mathbf{x_{2}}=f_{2}(\mathbf{x_{1}})$
is not continuous.

The above conjecture is demonstrated using the asymmetrically coupled
Lorenz system as an example. The equations for the asymmetrically
coupled Lorenz system\cite{belykh2006synchronization} are given by
\begin{align*}
\dot{x_{1}} & =\sigma(y_{1}-x_{1})+\alpha(x_{2}-x_{1}),\quad\dot{x_{2}}=\sigma(y_{2}-x_{2})\\
\dot{y_{1}} & =rx_{1}-y_{1}-x_{1}z_{1},\quad\dot{y_{2}}=rx_{2}-y_{2}-x_{2}z_{2}\\
\dot{z_{1}} & =x_{1}y_{1}-\beta z_{1},\quad\dot{z_{2}}=x_{2}y_{2}-\beta z_{2}
\end{align*}
 where $\sigma=10,~r=28,~\beta=8.0/3$ and $\alpha$ is the strength
of coupling. 
The above equations represent two different Lorenz systems with the
second system driving the first system. The first system is coupled
to the second system with the $x_{2}$ variable driving the dynamics
of $x_{1}$ via the coupling term $\alpha(x_{2}-x_{1})$. This coupling
is asymmetric as the dynamics of the second system represented by
$x_{2},~y_{2}$ and $z_{2}$ is independent of the first system represented
by variable $x_{1},~y_{1}$ and $z_{1}$.

This asymmetric coupling is demonstrated in Fig.~\ref{fig:coupledlorenz_x1x2_cstat}
and \ref{fig:coupledlorenz_x2x1_cstat} which shows the values of $\theta_{avg}(\epsilon_{f})$
for various values of coupling strength $\alpha$ for maps $\mathbf{x_{1}}=f_{1}(\mathbf{x_{2}})$
and $\mathbf{x_{2}}=f_{2}(\mathbf{x_{1}})$. The values of $\theta_{avg}(\epsilon_{f})$
for $f_{1}$ and $f_{2}$ for values of 1, 2, 4, 5, 7, 9 and 16 of $\alpha$
are shown. There is almost no coupling for $\alpha=1$ which is reflected
in low values in both directions, although even for this case it is
higher for $f_{1}$. As conjectured above, the directionality of coupling
can easily be seen for values of $\alpha$ of 2, 4, 5 and 7. The systems
become fully synchronized for large values of $\alpha$ which manifests
as full coupling in both direction as reflected for $\alpha$ of 9
and 16. Transfer entropy is unable to differentiate between no coupling
and fully synchronized systems\cite{Staniek} as both cases give a
value of zero. The present approach clearly differentiates between
no coupling and fully synchronized systems. No coupling is manifest
as nearly zero values of $\theta_{avg}(\epsilon_{f})$ in both direction.
Fully synchronized systems manifest as nearly a value of one for $\theta_{avg}(\epsilon_{f})$
in both direction.

\section{\label{sec:realworld}Complex time series from the finance world}

\begin{figure*}
    \centering
    \begin{subfigure}[t]{0.45\textwidth}
        \centering
        \includegraphics[width=\textwidth]{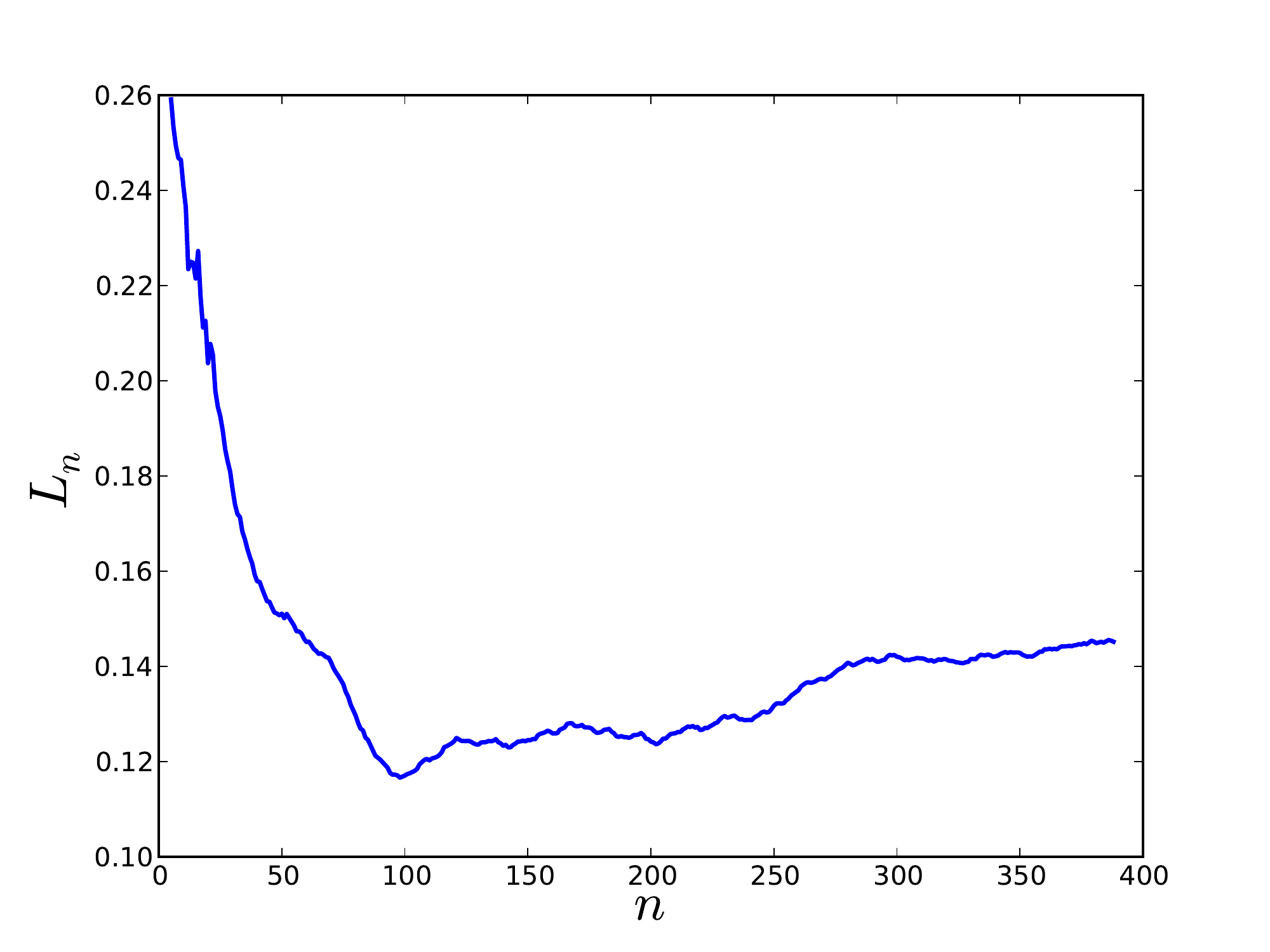} 
        \caption{Variation of $L_{n}$ for currency exchange
series. The minimum value is obtained for $n=98$ which is taken as
Markov order in the state space reconstruction process.}
        \label{fig:lncurrency}
    \end{subfigure}
    \begin{subfigure}[t]{0.45\textwidth}
        \centering
        \includegraphics[width=\textwidth]{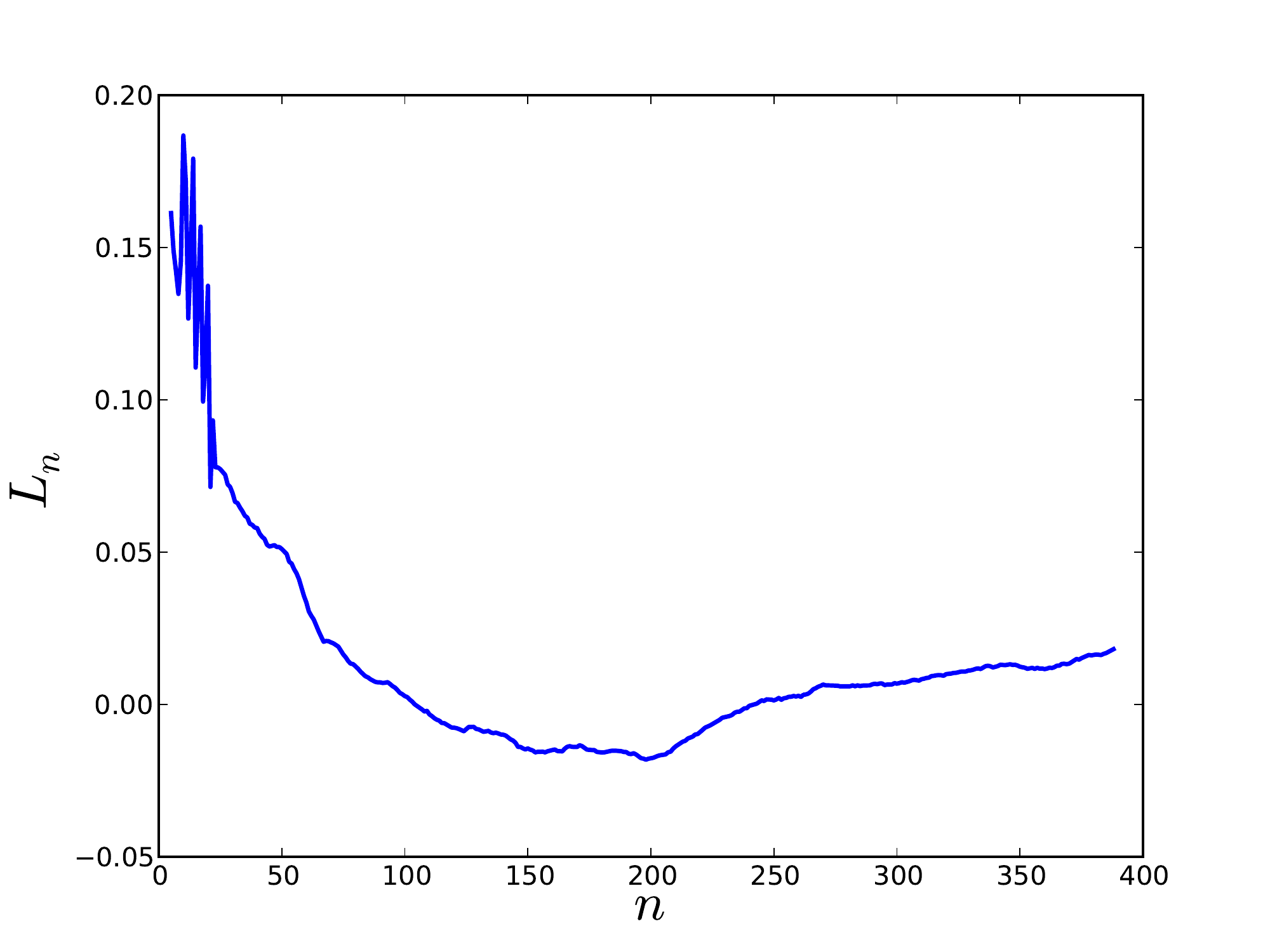} 
        \caption{Variation of $L_{n}$ for oil price series. The
minimum value is obtained for $n=198$ which is taken as Markov order
in the state space reconstruction process.}
        \label{fig:lnoil}
    \end{subfigure}
    \caption{Markov order for exchange rate and oil price fluctuation dynamics.}
\end{figure*}

In this section, the currency exchange rate Canadian Dollar-Japanese
Yen and oil prices between 2000 to 2014 are probed for coupling. 
The coupling between the Canadian Dollar-Japanese Yen\cite{canjpy}
exchange rate and the WTI crude oil price\cite{wti} is considered. 
\begin{figure}
\includegraphics[width=0.8\textwidth,height=0.4\textwidth]{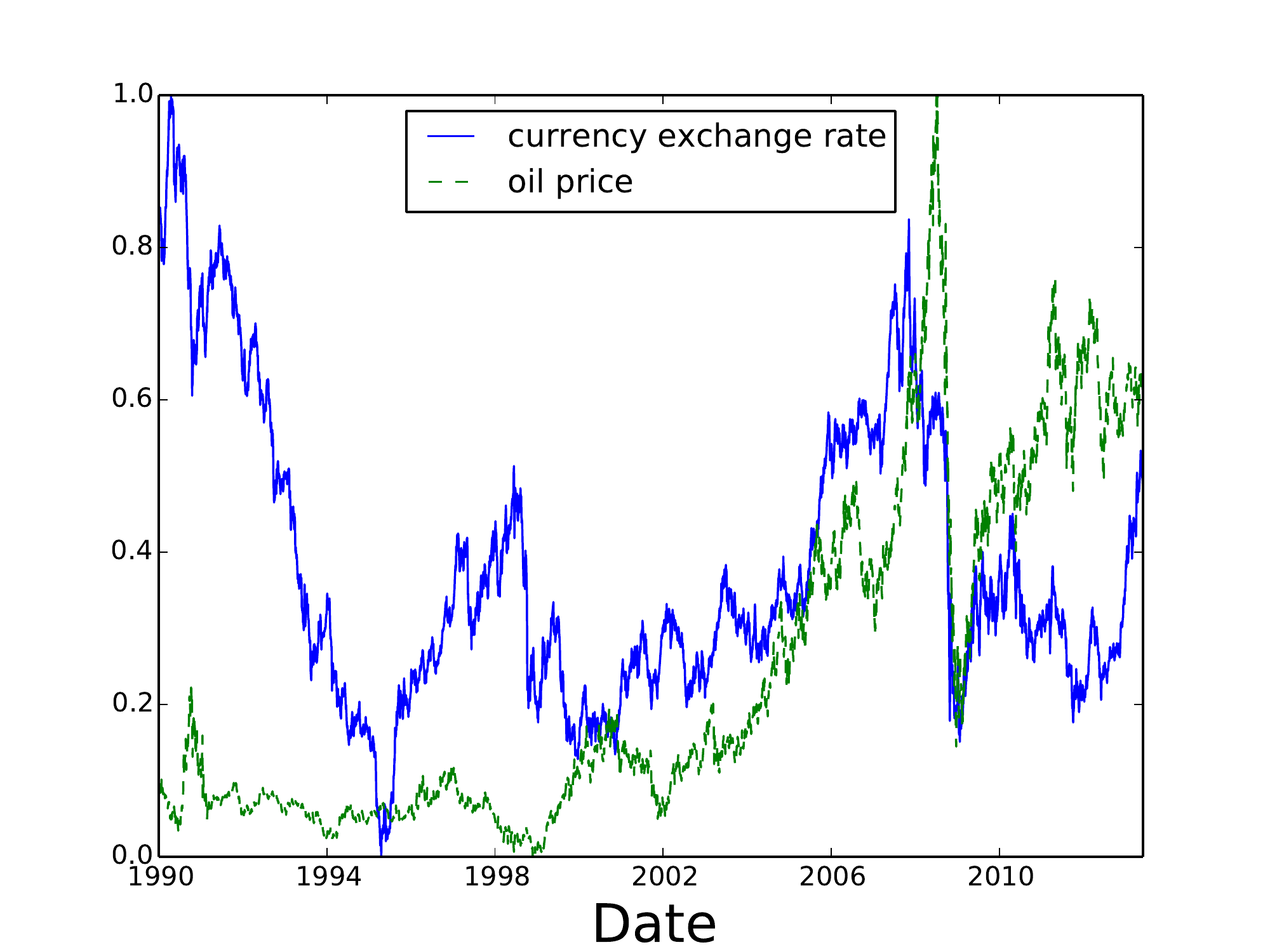} \caption{\label{fig:currencyoil}Currency exchange rate Canadian Dollar-Japanese
Yen and oil price between 1990 to 2013. The values have been rescaled
between 0 and 1.}
\end{figure}
\begin{figure*}
    \centering
    \begin{subfigure}[t]{0.45\textwidth}
        \centering
        \includegraphics[width=\textwidth]{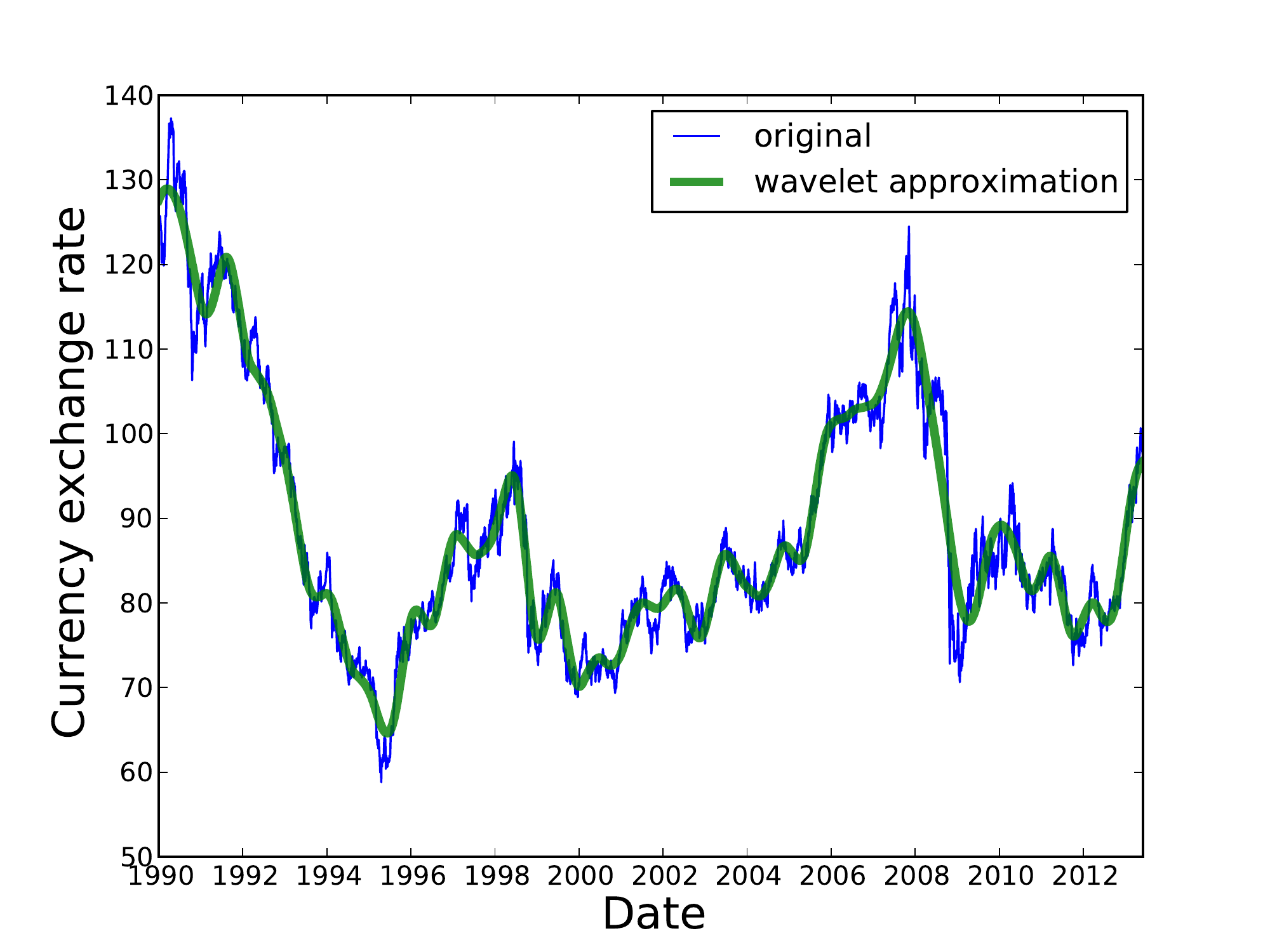} 
        \caption{Currency exchange rate Canadian Dollar-Japanese
Yen and oil price between 1990 to 2013. A level 7 approximation with
Daubechies wavelet filter is also shown superposed. This is subtracted
from the original time series in order to analyze the local fluctuation
dynamics.}
        \label{fig:currency_apprx}
    \end{subfigure}
    \begin{subfigure}[t]{0.45\textwidth}
        \centering
        \includegraphics[width=\textwidth]{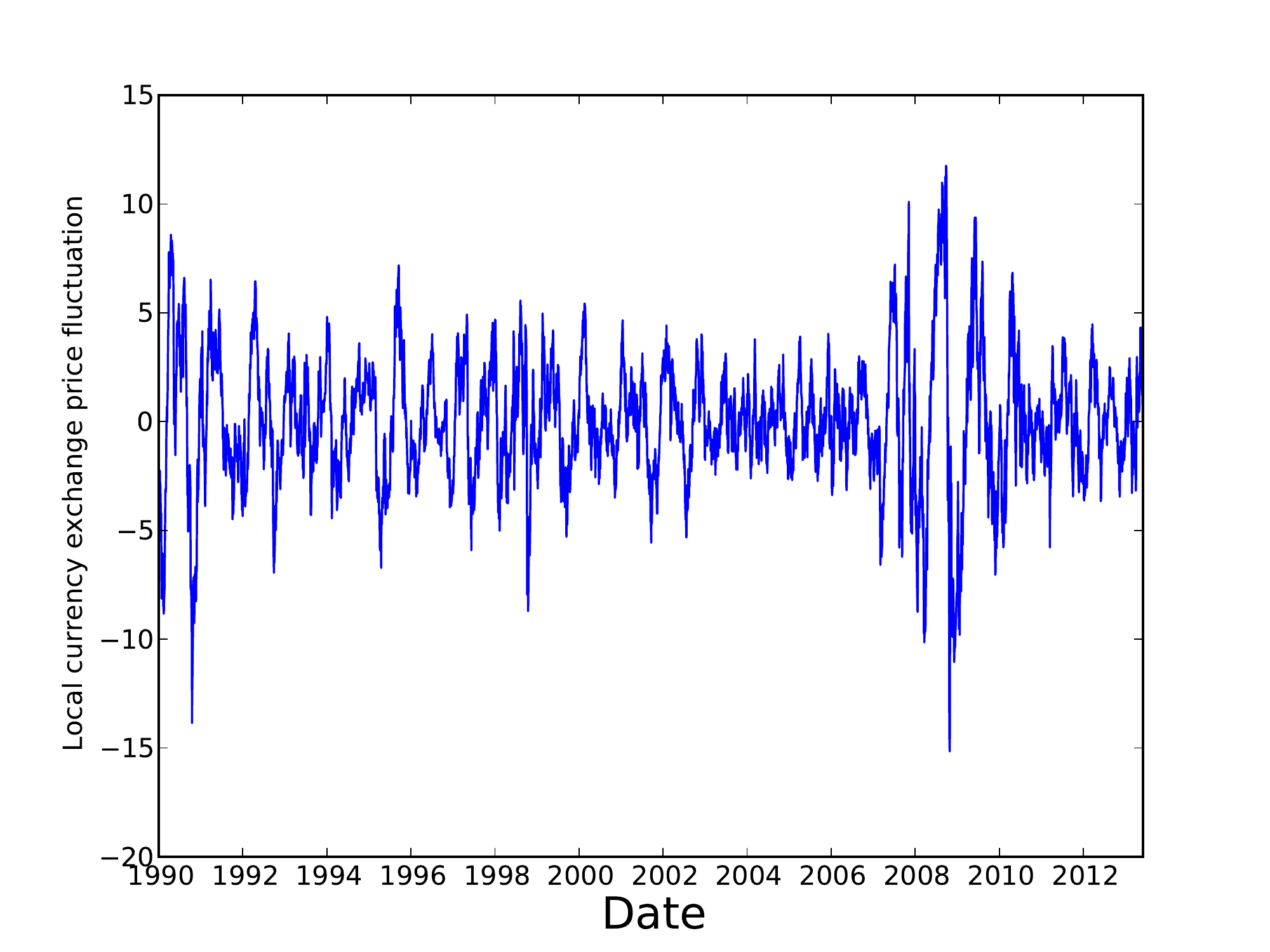} 
        \caption{Currency exchange rate local fluctuations after subtracting the wavelet approximation in Fig.~\ref{fig:currency_apprx}.}
        \label{fig:currency_localfluc}
    \end{subfigure}
    \caption{Exchange rate detrending using wavelet filter.}
\end{figure*}
\setcounter{subfigure}{0}
These variables are shown in Fig.~\ref{fig:currencyoil} for the
period 1990 to 2013. The values have been rescaled to take a value
between 0 and 1. It is hard to see any evidence of coupling by mere
visual inspection. 
In order to analyze both series for coupling, the series is detrended
by using Daubechies wavelet filter\cite{daubechies1990wavelet}. The
level 7 approximation superimposed on the original time series for
currency exchange rate is shown in Fig.~\ref{fig:currency_apprx}.
This approximation is subtracted from the series and local fluctuation
dynamics is probed for coupling. The local fluctuation for currency
exchange rate is shown in Fig.~\ref{fig:currency_localfluc}. The
same procedure is applied to oil price time series. The variation
$L_{n}$ with $n$ for currency exchange rate (CER) series and oil
price (OP) series is shown in Fig.~\ref{fig:lncurrency} and \ref{fig:lnoil}
respectively. Markov order of 98 and 198 are found for CER and OP
series. 
\setcounter{subfigure}{0}
\begin{figure}
\includegraphics[width=0.8\textwidth,height=0.4\textwidth]{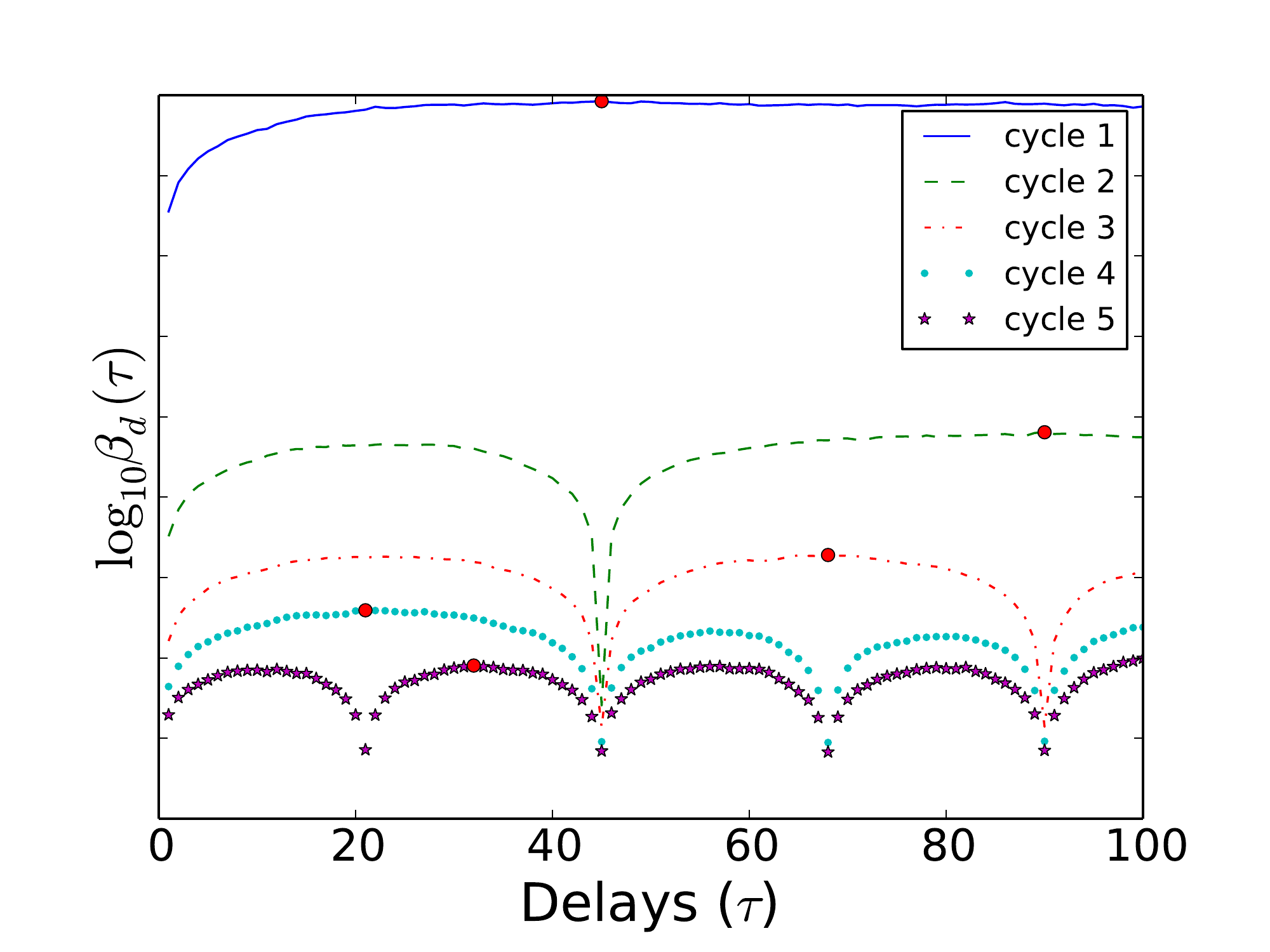} \caption{\label{fig:currency_recr}State space reconstruction for currency
exchange rate local fluctuation dynamics. Delays of 45, 90, 68, 21
and 32 are found to be optimal.}
\end{figure}
MDOP reconstruction methodology is then applied to these local fluctuation
series. The reconstruction for CER is shown in Fig.~\ref{fig:currency_recr}.
Delays of 45, 90, 68, 21 and 32 are found to be optimal. Using the
same procedure the delays of 155, 190, 37, 64 and 123 were found to be
optimal for the oil price fluctuation series. 
\begin{figure*}
    \centering
    \begin{subfigure}[t]{0.45\textwidth}
        \centering
        \includegraphics[width=\textwidth]{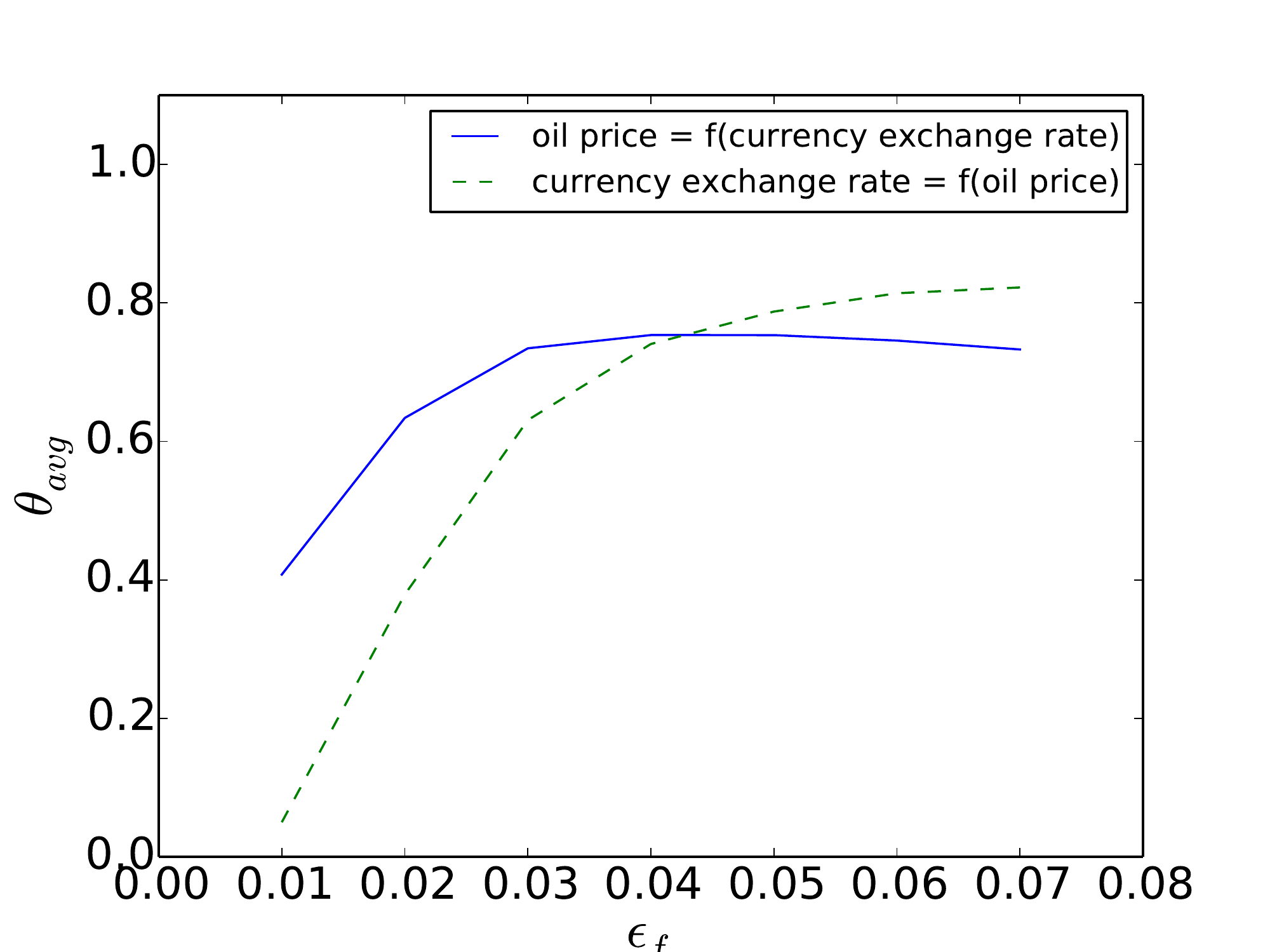} 
        \caption{$\theta_{avg}(\epsilon_{f})$ for probing
coupling between Canadian Dollar-Japanese Yen exchange rate and oil
price using minimal embedding technique. Strong coupling is found
between these two variables.}
        \label{fig:curroilcstat}
    \end{subfigure}
    \begin{subfigure}[t]{0.45\textwidth}
        \centering
        \includegraphics[width=\textwidth]{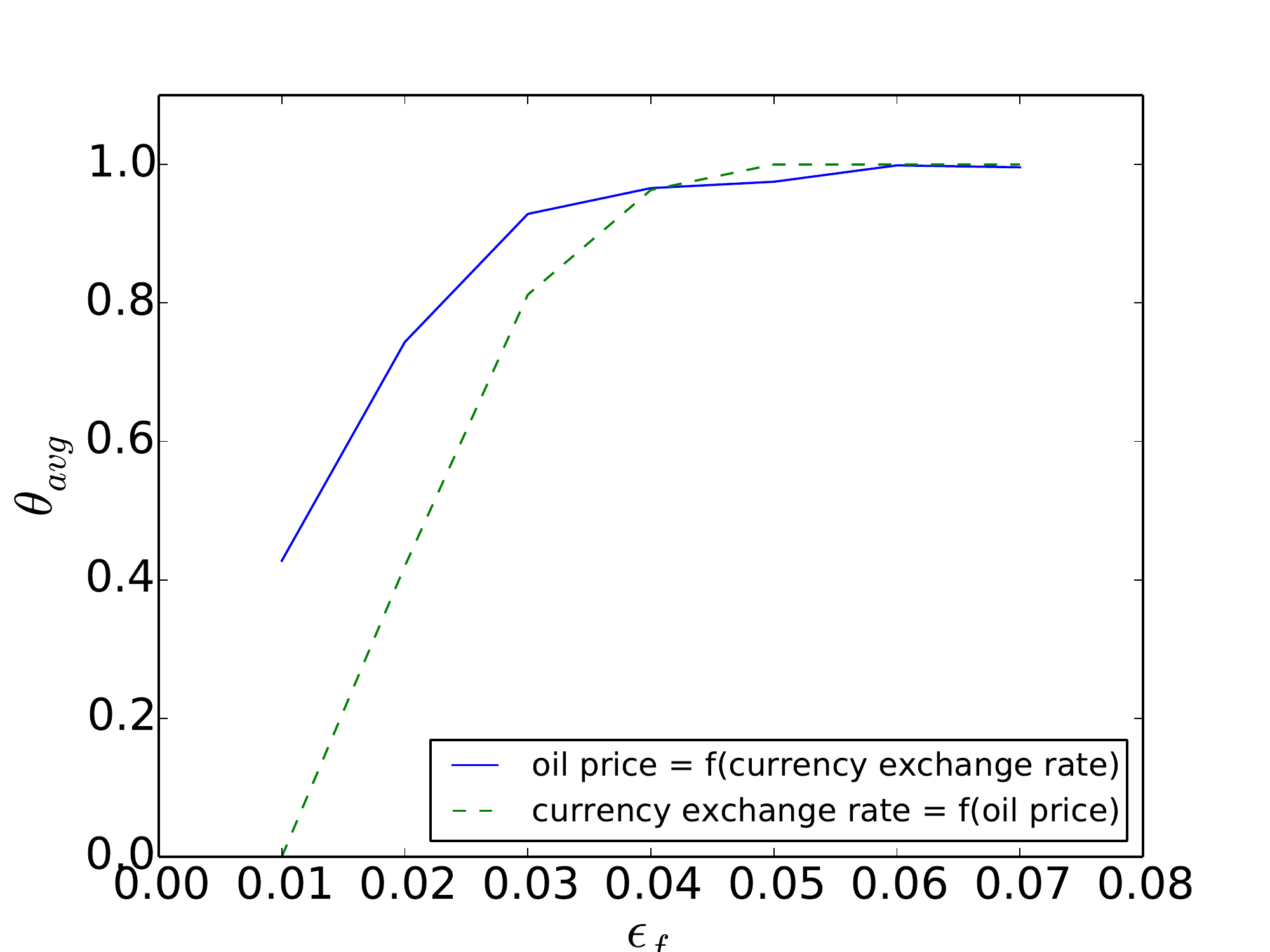} 
        \caption{$\theta_{avg}(\epsilon_{f})$ for probing
coupling between Canadian Dollar-Japanese Yen exchange rate and oil
price using very high embedding dimension. The measure is seen to
be convergent to a value of one indicating strong coupling.}
        \label{fig:curroilfullcstat}
    \end{subfigure}
    \caption{Coupling between exchange rate and oil fluctuation dynamics.}
\end{figure*}
\setcounter{subfigure}{0}
The values of $\theta_{avg}(\epsilon_{f})$ for maps between these
embeddings for values of $\epsilon_{f}$ ranging from 0 to 0.07 is
shown in Fig.~\ref{fig:curroilcstat}. 
The high dimensional embeddings for both series are also considered.
Fig.~\ref{fig:curroilfullcstat} shows the values of $\theta_{avg}(\epsilon_{f})$
when embedding dimension equal to Markov order with a time delay of one is used. 
The time series considered in this case is noisy, and, as described earlier in Section~\ref{sec:mdop},
the nearest neighbor ranks become more robust for higher embedding
dimension. Since the $\theta_{avg}(\epsilon_{f})$ measure depends
only on nearest neighbor rank and not on actual nearest neighbor distance,
it is more suitable to use high embedding dimension for noisy cases.
This of course comes at the penalty of computational cost because
nearest neighbor search is $\mathcal{O}(N^{2})$ operation for dimensions
greater than 20 even with fast search algorithms. Strong coupling
is evident from values in Figs.~\ref{fig:curroilcstat} and \ref{fig:curroilfullcstat}.
This example shows the inter-connectedness of economic forces in a
complex network.

\section{\label{sec:conclusions}Conclusions}

A physics based measure of coupling has been proposed. The measure
assesses the continuity of functional maps between time series embeddings
of two variables. It has been shown that it is necessary for the existence
of a continuous functional map between suitable time series embeddings
of two variables in order for coupling to be established. A mathematical
proof for this necessity has been provided. A statistic for continuity,
based on first principles definition of continuity, is used to probe
for coupling. This measure of continuity of the functional map is
convergent to a value of one in the case of coupling and zero in case
of no coupling. The proposed approach has been demonstrated by establishing
coupling between $x$ and $y$ variables of the Rossler system. The
measure is shown to be robust even in the presence of large amount
of observational noise. It has also been shown that the proposed approach
can also be used to assess the directionality of coupling. The directionality
can be determined if the functional map is found to be continuous
in one direction and not continuous in other direction. This was demonstrated
using unidirectionally coupled Lorenz oscillators. The measure can
also distinguish between fully synchronized oscillators and uncoupled
systems unlike the information-theoretic measure of transfer entropy
which is unable to distinguish between these two cases. The approach
is model-free and works very well with high-dimensional signals such as those found in financial settings. 
Density estimates in transfer entropy is an intractable problem in high dimensions.
The methodology presented does not require high-dimensional density
estimation. Two disparate economic variables, currency exchange rate
between the Canadian Dollar-Japanese Yen and oil prices have been
shown to be strongly coupled using this measure. 

\appendix

\section{\label{sec:CompositionProof}Composition of two continuous function
is a continuous function}

\textit{Definition}: A function $F(\mathbf{x})$ is continuous at
$\mathbf{x_{c}}$ if for every $\epsilon>0$ there exists a $\delta>0$
such that for all $\mathbf{x}$: 
\begin{align}
\left|\mathbf{x}-\mathbf{x}_{c}\right| & <\delta\Rightarrow\left|F(\mathbf{x})-F(\mathbf{x}_{c})\right|<\epsilon~.\label{eq:fcontinuity}
\end{align}

\textit{Proposition}: If two function $F_{1}(\mathbf{x})$ and $F_{2}(\mathbf{x})$
are continuous at $\mathbf{x_{c}}$ then their composition $(F_{1} \circ F_{2})(\mathbf{x})$
is also continuous at $\mathbf{x_{c}}$.

\textit{Proof}: Let $F_{1}$ and $F_{2}$ be two continuous functions.
Consider a point $\mathbf{x}_{c}$. Since $F_{1}$ is continuous,
for every $\epsilon>0$ there exists a $\eta>0$ such that for all
$\mathbf{x}$: 
\begin{align}
\left|F_{2}(\mathbf{x})-F_{2}(\mathbf{x}_{c})\right| & <\eta\Rightarrow\left|(F_{1} \circ F_{2})(\mathbf{x})-(F_{1} \circ F_{2})(\mathbf{x}_{c})\right|<\epsilon~.\label{eq:f1f2continuity}
\end{align}
 Since $F_{2}$ is continuous, for every $\eta>0$ there exists a
$\delta>0$ such that for all $\mathbf{x}$: 
\begin{align}
\left|\mathbf{x}-\mathbf{x}_{c}\right| & <\delta\Rightarrow\left|F_{2}(\mathbf{x})-F_{2}(\mathbf{x}_{c})\right|<\eta~.\label{eq:f2continuity}
\end{align}
 It follows from assertions~(\ref{eq:f1f2continuity}) and (\ref{eq:f2continuity})
that for every $\epsilon>0$ there exists a $\delta>0$ such that
for all $\mathbf{x}$: 
\begin{align*}
\left|\mathbf{x}-\mathbf{x}_{c}\right| & <\delta\Rightarrow\left|(F_{1} \circ F_{2})(\mathbf{x})-(F_{1} \circ F_{2})(\mathbf{x}_{c})\right|<\epsilon~.
\end{align*}
 It thus follows from definition~(\ref{eq:fcontinuity}) that the
composition function $(F_{1} \circ F_{2})(\mathbf{x})$ is continuous at
$\mathbf{x}_{c}$.

\begin{acknowledgments}
This research was supported by the Australian Research Council and Sirca Technology Pty Ltd under Linkage Project LP100100312.
The author is also supported by International Macquarie University Research Excellence Scholarship (iMQRES).
Deb Kane is thanked for her editorial support and critical reading of the manuscript.
\end{acknowledgments}

\bibliography{pre5}

\end{document}